\documentclass[12pt]{JHEP3}

 \preprint{  }

\usepackage{epsfig,multicol}
\usepackage{amsmath}
\usepackage{graphics}
\usepackage{graphicx}
\usepackage{dcolumn}
\usepackage{amssymb}

\title{Holographic Superconductors with Power-Maxwell field}
\author{
Jiliang  Jing\footnote{Email:
jljing@hunnu.edu.cn}$^{1}$, Qiyuan Pan, Songbai Chen
\\ Institute of Physics and
Department of Physics, and Key Laboratory of Low Dimensional Quantum Structures and
Quantum Control of Ministry of Education, Hunan Normal University,
Changsha, Hunan 410081, P. R. China}

\abstract{
With the Sturm-Liouville analytical and numerical methods, we investigate the behaviors of the holographic superconductors by introducing a complex charged scalar field coupled with a Power-Maxwell field in the background of $d$-dimensional Schwarzschild AdS black hole. We note that the Power-Maxwell field takes the special asymptotical solution near boundary which is different from all known cases. We find that the larger power parameter $q$ for the Power-Maxwell field makes it harder for the scalar hair to be condensated. We also find that, for different $q$, the critical exponent of the system is still $1/2$, which seems to be an universal property for various nonlinear electrodynamics if the scalar field takes the form of this paper.}


\begin{document}

\section{Introduction}

The AdS/CFT correspondence \cite{Maldacena,polyakov,Witten}
indicates that a weak coupling gravity theory in a $d$-dimensional
anti-de Sitter spacetime can be related to a strong coupling
conformal field theory on the $(d-1)$-dimensional boundary. Gubser
first \cite{Gubser:2005ih,GubserPRD78} suggested that near the
horizon of a charged black hole there is in operation a geometrical
mechanism parameterized by a charged scalar field of breaking a
local $U(1)$ gauge symmetry. Then, the gravitational dual of the
transition from normal to superconducting states in the boundary
theory was constructed. This dual consists of a system with a black
hole and a charged scalar field, in which the black hole admits
scalar hair at temperature lower than a critical temperature, but
does not possess scalar hair at higher temperatures
\cite{HartnollPRL101}. Since the AdS/CFT duality is a valuable tool
for investigating strongly coupled gauge theories, the application
might offer new insight into the investigation of strongly
interacting condensed matter systems where the perturbational
methods are no longer available. Therefore, much attention has been
given to the studies of the AdS/CFT duality to condensed matter
physics and in particular to superconductivity
\cite{HartnollJHEP12,HorowitzPRD78,Nakano-Wen,Amado,
Koutsoumbas,Maeda79,Sonner,HartnollRev,HerzogRev,
Ammon:2008fc,Gubser:2009qm,CJ0,Gregory,Pan-Wang,
Ge-Wang,Brihaye,Gregory1009,Pan-Wang1,Cai-pGB,Jing-Chen} recently.

Because the Maxwell theory is only a special case or a leading order
in the expanded form of nonlinear electrodynamics, the nonlinear
electrodynamics which carries more information than the Maxwell
field has been a subject of research for many years
\cite{euler,born,gibb,Olivera1, Hoffman-Gibbons-Rasheed, Oliveira}.
Heisenberg and Euler \cite{euler} noted that quantum electrodynamics
predicts that the electromagnetic field behaves nonlinearly through
the presence of virtual charged particles. Now we list the main
nonlinear electrodynamics as follows: (i) Born and Infeld
\cite{born} presented a classical nonlinear theory of
electromagnetism which contains many symmetries common to the
Maxwell theory despite its nonlinearity. The Lagrangian density for
Born-Infeld theory is $\mathcal{L_{BI}}=4b^2 \left(1-\sqrt{1
+\frac{F^2}{2b^2}}\right)$ with $F^2=F_{\mu\nu}F^{\mu\nu}$ and the
coupling parameter $b$ is related to the string tension
$\alpha^{\prime}$ as $b =1/(2\pi \alpha^{\prime})$. This Lagrangian
reduces to the Maxwell case in the weak-coupling limit
$b\rightarrow\infty $. (ii) The action of Power-Maxwell filed
\cite{HM,HMM,MC,Olivera1,Gurtug} is taken as power-law function of
the form $\mathcal{L_{BI}}=- \beta F^{2 q}$, where $\beta$ is a
coupling constant and $q$ is a power parameter. It is interesting
that the conformal invariance $g_{\mu\nu}\rightarrow \Omega^2
g_{\mu\nu}$, $A_{\mu}\rightarrow A_{\mu}$ is realized for the power
parameter $q=d/4$ where $d$ is the dimensions of the spacetime
\cite{HM}. (iii) The nonlinear electromagnetic Lagrangian that
contains logarithmic terms appears in the description of vacuum
polarization effects. The term were obtained as exact 1-loop
corrections for electrons in a uniform electromagnetic field
background by Euler and Heisenberg \cite{euler}. A simple example of
a Born-Infeld-like Lagrangian with a logarithmic term, that can be
added as a correction to the original Born-Infeld one, was discussed
in Ref. \cite{HH}. In an arbitrary dimension, the logarithmic
electromagnetic lagrangian has the form $\mathcal{L_{BI}}=- b^2 \ln
\left( 1+\frac{F^2}{b^2}\right)$ where $b$ is a coupling constant.
The Lagrangian tends to the Maxwell case in the weak-coupling limit
$b\rightarrow \infty $.

It is well known that the properties of holographic superconductors
depend on behavior of the electromagnetic field coupled with the
charged scalar filed in the system. Motivated by the recent studies
and the fact that, within the framework of AdS/CFT correspondence,
the different electromagnetic action is expected to modify the
dynamics of the dual theory, in this paper we will investigate the
behavior of the holographic superconductors with the Power-Maxwell
field in the background of a $d$-dimensional Schwarzschild AdS black
hole, and to see how the Power-Maxwell field affect the formation of
the scalar hair and the critical exponent of the system.

The paper is organized as follows. In Sec. II,  we give the
holographic dual of $d$-dimensional Schwarzschild AdS black hole by
introducing a complex charged scalar field coupled with a
Power-Maxwell field. In Sec. III, we explore the relations between
critical temperature and charge density. In Sec. IV, we study the
critical exponents of the holographic superconductor model with the
Power-Maxwell field. We summarize and discuss our conclusions in the
last section.

\section{Holographic dual of $d$-dimensional Schwarzschild AdS
black hole}

In order to study a superconductor dual to a AdS black hole configuration in the probe
limit, we consider the $d$-dimensional Schwarzschild AdS black hole
\begin{eqnarray}\label{BH metric}
ds^2=-f(r)dt^{2}+\frac{dr^2}{f(r)}+r^{2}dx_{i}dx^{i},
\end{eqnarray}
with
\begin{eqnarray}
f(r)=r^2\left(1-\frac{r_+^{d-1}}{r^{d-1}}\right),
\end{eqnarray}
where we have chosen units such that the AdS radius is
unity, and $r_+$ is radius of the event horizon. The Hawking temperature
of the black hole is
\begin{eqnarray}
\label{Hawking temperature} T=\frac{(d-1)r_{+}}{4\pi}\ .
\end{eqnarray}

We now consider the Power-Maxwell field and the charged  scalar
field coupled via a Lagrangian
\begin{eqnarray}\label{System}
S=\int d^{d}x\sqrt{-g}\left[
-\beta \left(F_{\mu\nu}F^{\mu\nu}\right)^q
-\partial_{\mu}\tilde{\psi}\partial^{\mu}\tilde{\psi}
-m^2\tilde{\psi}^2-\tilde{\psi}^2(\partial_{\mu}p-A_{\mu})
(\partial^{\mu}p-A^{\mu}) \right] \ ,\nonumber\\
\end{eqnarray}
where $F^{\mu\nu}$ is the strength of the Power-Maxwell field $F=dA$
and $\tilde{\psi}$ is the complex scalar field, and $\beta$ and $q$
are the coupling constant and the power parameter of the
Power-Maxwell field, respectively.  The Power-Maxwell field will
reduce to the Maxwell case when both $\beta=1/4$ and $q=1$. We can
use the gauge freedom to fix $p=0$ and take
$\psi\equiv\tilde{\psi}$, $A_{t}=\phi$ where $\psi$, $\phi$ are both
real functions  of $r$ only. Then  the equations of motion are given
by
\begin{eqnarray}\label{Psi}
&&\psi^{\prime\prime}+\left(
\frac{f^\prime}{f}+\frac{d-2}{r}\right)\psi^\prime
+\frac{\phi^2}{f^2}\psi-\frac{m^2}{f}\psi=0\,,
\\ \label{Phi} &&
\phi^{\prime\prime}+\left(\frac{d-2}{2q-1}
\right)\frac{\phi^\prime}{r} -\frac{2}{(-2)^{1+q}\beta q (2q-1)}
\frac{\psi^2\phi}{f}(\phi^\prime)^{2(1-q)}=0,
\end{eqnarray}
where  a prime denotes the derivative with respect to $r$. At the
event horizon $r=r_+$, we must have$^{1}$ \footnotetext[1]{From Eq.
(\ref{Phi}) we know that at the event horizon there may be two
cases, i.e., $\phi(r_{+})=0$ or $\phi^\prime(r_{+})=0$. The case of
$\phi^\prime(r_{+})=0$ shows that $\phi(r_{+})=constant$. However,
the constant must be set to zero according to Gubser's argument in
Ref. \cite{GubserPRD78}.}
\begin{eqnarray}\label{Nearh}
 &&\phi(r_{+})=0,\nonumber \\
 &&\psi(r_{+})=\frac{(d-1)r_+ }{ m^{2}}\psi^\prime(r_{+}),
\end{eqnarray} and at the asymptotic AdS region
($r\rightarrow\infty$), the solutions behave like ( For behavior of $\phi$ please see Appendix  for detail)
\begin{eqnarray}
&&\psi=\frac{\psi_{-}}{r^{\lambda_{-}}}+
\frac{\psi_{+}}{r^{\lambda_{+}}}, \nonumber \\
&& \phi=\mu-\frac{\rho^\frac{1}{2q-1}}{r^{\frac{d-2}{2q-1}-1}}, \label{infinity}
\end{eqnarray}
with
\begin{eqnarray}
&&\lambda_\pm=\frac{1}{2}\left[(d-1)\pm\sqrt{(d-1)^{2}
+4m^{2}}~\right], \label{LambdaZF}
\end{eqnarray}
where $\mu$ and $\rho$ are interpreted as the chemical potential and charge density in the dual field theory, respectively. The
coefficients $\psi_+$ and $\psi_-$ both multiply normalizable modes
of the scalar field equations and they correspond to the vacuum
expectation values $\psi_+=\langle \mathcal{O_+}\rangle$,
$\psi_-=\langle \mathcal{O_-}\rangle$ of an operator $\mathcal{O}$
dual to the scalar field according to the AdS/CFT correspondence. We can impose boundary conditions that either $\psi_+$ or $\psi_-$
vanishes. It is of interest to note that the electric field $\phi$
is dependent on the coupling constant $q$ of the Power-Maxwell field at the asymptotic AdS region, which is different from the
Born-Infeld electrodynamics \cite{Jing} and all known cases.

\section{Relations between critical temperature and charge density}

In this section we will use both the Sturm-Liouville analytical
\cite{Fz} and numerical methods to calculate the relation between
the critical temperature and the charge density  of the holographic
superconductors in Schwarzschild AdS black hole with the
Power-Maxwell field.

Introducing a new coordinate $z=\frac{r_+}{r}$, we can rewrite Eqs.
(\ref{Psi}) and  (\ref{Phi}) as
\begin{eqnarray}
\label{Psiz}
&&\psi^{\prime\prime}+\left(
\frac{f^\prime}{f}-\frac{d-4}{z}\right)\psi^\prime
+\frac{r_+^2}{z^4}\left(\frac{\phi^2}{f^2}-\frac{m^2}{f}\right)\psi=0\,,
\\ \label{Phiz} &&
\phi^{\prime\prime}-\frac{1}{z}\left(\frac{d-2}{2q-1}-2\right)\phi^\prime
-\frac{2 r_+^{2 q}}{(-1)^{1+3q}2^{1+q}\beta q (2q-1) z^{4 q}}\frac{\psi^2 \phi}{f}(\phi^\prime)^{2(1-q)}=0,
\end{eqnarray}
here and hereafter a prime denotes the derivative with respect to $z$.
When the temperature $T$ approaches the critical temperature $T_c$, the condensation approaches zero, viz. $\psi\rightarrow 0$. Thus, Eq. (\ref{Phiz}) becomes
\begin{eqnarray}
\label{Phiz0}
\phi^{\prime\prime}-\frac{1}{z}\left(\frac{d-2}{2q-1}-2\right)\phi^\prime
\approx 0.
\end{eqnarray}
The general solution of this equation takes form $\phi=\tilde{a}+\tilde{b}
z^{\frac{d-2}{2q-1}-1}$. With help of the boundary conditions
(\ref{Nearh}) and (\ref{infinity}),  we find that, near the critical
temperature, the electric field can be expressed as
\begin{eqnarray}\label{phitc}
\phi=\xi r_+\left(1- z^{\frac{d-2}{2q-1}-1}\right),
\end{eqnarray}
where $\xi=\left(\frac{\rho}{r_+^{d-2}}\right)^\frac{1}{2q-1}$.

Near the event horizon, we introduce a trial  function $F(z)$ into
$\psi$ as in Ref. \cite{Fz}
\begin{eqnarray}\label{OO}
\psi|_{z=0}\sim \frac{\psi_i}{r^{\lambda_i}} \sim
\langle\mathcal{O}_i
\rangle\frac{z^{\lambda_i}}{r_+^{\lambda_i}}F(z),
\end{eqnarray}
here and hereafter subscript $i=(+,~-)$. The trial  function should
satisfy $F(0)=1$ and $F^\prime(0)=0$. Then, using Eqs. (\ref{phitc})
and (\ref{OO}), the equation (\ref{Psiz}) for $\psi$ can be
rewritten as
\begin{eqnarray}\label{psiFz}
F^{\prime\prime}(z)&+&\left\{\frac{2\lambda_i}{z}
 -\left[\frac{2+(d-3)z^{d-1}} {z(1-z^{d-1})}
 +\frac{d-4}{z}\right]\right\}F^\prime(z)\nonumber \\
&+&\left\{\frac{\lambda_i(\lambda_i-1)}{z^2}- \frac{\lambda_i}{z}
\left[\frac{2+(d-3)z^{d-1}}{z(1-z^{d-1})}+ \frac{d-4}{z}\right]
\right. \nonumber \\ &+&\left.\frac{1}{1-z^{d-1}}
\left[\frac{\xi^2(1-z^{\frac{d-2}{2q-1}-1})^2}{1-z^{d-1}}
-\frac{m^2}{z^2}\right]\right\}F(z)=0.
\end{eqnarray}
Multiplying the above equation with the following functional
\begin{eqnarray}
T(z)=z^{2\lambda_i-d+2}(z^{d-1}-1),
\end{eqnarray}
we can express Eq. (\ref{psiFz}) as
\begin{eqnarray}\label{psiT}
[T(z) F^\prime(z)]^\prime-Q(z)F(z)+\xi^2 P(z) F(z)=0,
\end{eqnarray}
with
\begin{eqnarray}
&&Q(z)=-T(z)\left\{\frac{\lambda_i(\lambda_i-1)}{z^2}-\frac{\lambda_i}{z} \left[\frac{2+(d-3)z^{d-1}}{z(1-z^{d-1})}+\frac{d-4}{z}\right]-\frac{m^2}{z^2(1-z^{d-1})} \right\},\nonumber \\
&&P(z)=T(z) \frac{(1-z^{\frac{d-2}{2q-1}-1})^2}{(1-z^{d-1})^2}.
\end{eqnarray}
By using Sturm-Liouville method \cite{Fz} to solve the Eq.
(\ref{psiT}), we know that the minimum of eigenvalues of $\xi^2$ can
be obtained from the variation of the following expression
\begin{eqnarray}\label{xi2}
\xi^2=\frac{\int_0^1 T(z)F^{\prime}(z)^2 d z  +\int_0^1 Q(z) F(z)^2
d z}{\int_0^1 P(z) F(z)^2 d z}.
\end{eqnarray}
The trial function $F(z)$ can be taken as $F(z)=1-az^2$ which
satisfies its boundary condition. Then $\xi^2$ can be explicitly
written as
\begin{eqnarray}\label{xi22}
\xi^2(d,q,m,a)=\frac{s(d,q,m,a)}{t(d,q,m,a)}.
\end{eqnarray}
For different values of $d$, $q$ and $m$, we can find the minimum value of $\xi^2$ with
appropriate value of $a$. For example, taking $d=4$ and $q=3/4$, we have
\begin{eqnarray}\label{xi23}
s(4,3/4,m,a)&=& -\frac{12 a^2}{4 m^2+11 \sqrt{4
   m^2+9}+37}-\frac{1}{2} \left(2 m^2+3 \sqrt{4
   m^2+9}+9\right)\nonumber\\
   &&\times \left(\frac{a^2}{\sqrt{4
   m^2+9}+7}-\frac{2 a}{\sqrt{4
   m^2+9}+5}+\frac{1}{\sqrt{4
   m^2+9}+3}\right),\nonumber \\
t(4,3/4,m,a)&=&-\frac{3 a^2}{4 m^2+15
   \sqrt{4 m^2+9}+63}+\frac{6 a}{4 m^2+11
   \sqrt{4 m^2+9}+37}\nonumber\\ &&-\frac{3}{4 m^2+7 \sqrt{4
   m^2+9}+19}.
\end{eqnarray}
From which we can obtain $\xi_{min}=3.96555$ with $m^2=0$ when
$a=0.71075$,   $\xi_{min}=3.30506$ with $m^2=-1$ when $a=0.63430$,
$\xi_{min}=2.28183$ with $m^2=-2$ when $a=0.47874$, and
$\xi_{min}=1.5073$ with $m^2=-9/4$ when $a=0.33117$.

With the help of $T=\frac{(d-1)r_+}{4\pi}$ and $\xi=\left(\frac{\rho}{r_+^{(d-2)}}\right)^\frac{1}{2q-1}$, we know that, when $T\sim T_c$, the critical
temperature $T_c$ can be expressed as
\begin{eqnarray}\label{Txi}
T_c= \gamma \rho^{\frac{1}{d-2}},
\end{eqnarray}
where the coefficient $\gamma=\frac{d-1}{4 \pi
\xi_{min}^{(2q-1)/(d-2)}}$.

In Tables \ref{Tc-a} and \ref{Tc-b}, we list the analytical values
and numerical values of critical temperature for different $q$ and
$m$ in the 4-dimensional and 5-dimensional black holes,
respectively. Some numerical data for critical temperature with
$q=1$ are taken from the Ref. \cite{HorowitzPRD78}. The differences
between the analytical and numerical values are within $5\%$.

{\scriptsize{ \TABLE{
\caption{\label{Tc-a} The critical values of $T_c$ for different $q$
and $m$ in 4-dimensional black hole. The numerical data for critical
temperature with $q=1$ are taken from the Ref. \cite{HorowitzPRD78}}
\begin{tabular}{c | c | c | c | c | c | c | c | c }   \hline
$m^2$&\multicolumn{4}{c|}{$ q=\frac{3}{4} $}&\multicolumn{4}{c} {$ q=1 $}  \\
         \hline
 &\multicolumn{2}{c|}{$ T_c$  for $ \lambda_+$}&\multicolumn{2}{c|} {$ T_c$ for $ \lambda_- $}&\multicolumn{2}{c|}{$ T_c$ for $\lambda_+ $}&\multicolumn{2}{c} {$ T_c$ for $ \lambda_- $}  \\
         \hline
&Analytical&Numerical&Analytical&Numerical&Analytical&Numerical&Analytical&Numerical\\          \hline
$ \ \ 0$&$0.1692\rho^{\frac{1}{2}}$&$0.1694\rho^{\frac{1}{2}}$&---& ---&$0.0844 \rho^{\frac{1}{2}}$&$0.0870\rho^{\frac{1}{2}}$&--- & ---
                 \\
\hline
$-2$&$0.1942\rho^{\frac{1}{2}}$&$0.1943\rho^{\frac{1}{2}}$&$0.2529\rho^{\frac{1}{2}}$ &$0.2528 \rho^{\frac{1}{2}}$
&$0.1170\rho^{\frac{1}{2}}$&$0.1180\rho^{\frac{1}{2}}$&$0.2250\rho^{\frac{1}{2}}$&
$0.2260\rho^{\frac{1}{2}}$
                 \\
\hline
$-\frac{9}{4}$ &$0.2155\rho^{\frac{1}{2}}$ & $0.2154\rho^{\frac{1}{2}}$&$0.2155\rho^{\frac{1}{2}}$ &$0.2154 \rho^{\frac{1}{2}}$
&$0.1507 \rho^{\frac{1}{2}}$&$0.1520\rho^{\frac{1}{2}}$&$0.1507\rho^{\frac{1}{2}}$&$0.1520 \rho^{\frac{1}{2}}$
                 \\
        \hline
\end{tabular}
}}}

{\footnotesize{ \TABLE{ \caption{\label{Tc-b} The critical values of $T_c$ for different $q$
and $m$ in 5-dimensional black hole. The some numerical data for
critical temperature with $q=1$ are taken from the Ref.
\cite{HorowitzPRD78}}
\begin{tabular}{c || c | c || c | c || c | c }   \hline
$m^2$ & \multicolumn{2}{c||}{$ q=\frac{3}{4} $} &\multicolumn{2}{c||} {$ q=1 $} &\multicolumn{2}{c}{$ q=\frac{5}{4} $} \\
         \hline
 & \multicolumn{2}{c||}{$ T_c$  for $ \lambda_+$} &
 \multicolumn{2}{c||}{$ T_c$ for $\lambda_+ $} & \multicolumn{2}{c} {$ T_c$ for $ \lambda_+ $}  \\
         \hline
 & Analytical &  Numerical
 & Analytical &  Numerical &  Analytical & Numerical
\\          \hline
$\ \  0$ &$0.2503 \rho^{\frac{1}{3}}$ &$0.2505 \rho^{\frac{1}{3}}$ & $0.1676 \rho^{\frac{1}{3}}$  &
$0.1700 \rho^{\frac{1}{3}}$& $0.0954\rho^{\frac{1}{3}}$ &$0.1008\rho^{\frac{1}{3}}$
                 \\
\hline $-1$ &$0.2543 \rho^{\frac{1}{3}}$ & $0.2545\rho^{\frac{1}{3}}$& $0.1739 \rho^{\frac{1}{3}}$
& $0.1765\rho^{\frac{1}{3}}$& $0.1014\rho^{\frac{1}{3}}$ &$0.1065\rho^{\frac{1}{3}}$
                 \\
\hline $-2$ &$0.2596 \rho^{\frac{1}{3}}$ & $0.2746\rho^{\frac{1}{3}}$& $0.1825
\rho^{\frac{1}{3}}$ & $0.1847  \rho^{\frac{1}{3}}$ &$0.1099 \rho^{\frac{1}{3}}$ & $0.1145
\rho^{\frac{1}{3}}$
                 \\
\hline $-3$ &$0.2677 \rho^{\frac{1}{3}}$ & $0.2642 \rho^{\frac{1}{3}}$ &
$0.1962 \rho^{\frac{1}{3}}$  & $0.1980 \rho^{\frac{1}{3}}$&$0.1240 \rho^{\frac{1}{3}}$  &
$0.1279 \rho^{\frac{1}{3}}$
                 \\
        \hline
\end{tabular}
}}}

From tables I and II, we find that, for the same $q$, the critical
temperature for the scalar operators $\langle\mathcal{O}_+ \rangle$
decreases as the value of $m^2$ increases, which means that the
larger mass ($m^2$ becomes less negative) of the scalar field makes
it harder for the scalar hair to be condensated in both
4-dimensional and 5-dimensional Schwarzschild AdS black holes, which
agrees with the finding in Ref. \cite{HorowitzPRD78}.

From the tables we also know that, for the same mass $m$ and
fixed scalar operators $\langle\mathcal{O}_i \rangle$ with
$i=(+,-)$, the ratio $T_{c}/\rho^{1/(d-2)}$ decreases as the $q$
increases, which means that the larger power $q$ for the
Power-Maxwell field makes it harder for the scalar hair to be
condensated in the Schwarzschild AdS black hole.

\section{Critical exponents}

In this section, we will study the critical exponents of the
holographic superconductor model with the Power-Maxwell field by
using analytical and numerical methods, respectively.

Now we are in position to investigate the critical exponents
analytically. From last section we know that the condensation value
of the dual operator $\langle\mathcal{O}_i \rangle$ is very small
when $T\rightarrow T_c$. Substituting Eq. (\ref{OO}) into Eq.
(\ref{Phiz}), we have

\begin{eqnarray}\label{phi}
\phi^{\prime\prime}-\frac{1}{z}\left(\frac{d-2}{2q-1}-2\right)\phi^\prime
=\frac{2r_+^{2 q-2\lambda_i-2} \langle\mathcal{O}_i \rangle^2
}{(-1)^{1+3q}2^{1+q}\beta q (2q-1) }\frac{z^{2\lambda_i-4 q} F^2(z)
\phi}{g(z)}(\phi^\prime)^{2(1-q)},
\end{eqnarray}
where $g(z)=(1-z^{d-1})/z^2$. Note that,  near the critical
temperature, Eq. (\ref{phitc}) can be rewritten as
\begin{eqnarray}\label{phin}
\phi=\frac{A T_c^{\frac{d-2}{2q-1}}}
{T^{\frac{d-2}{2q-1}-1}}\left(1-z^{\frac{d-2}{2q-1}-1}\right),
\end{eqnarray}
where $A=\frac{d-1}{4\pi} $. Therefore, we can assume that the
general solution for Eq. (\ref{phi}) takes the form
\begin{eqnarray}\label{phis}
\phi=A T_c(1-z^{\frac{d-2}{2q-1}-1})+\left(A T_c
\right)^m\left[\frac{r_+^{2 q-2 \lambda_i-2} \langle\mathcal{O}_i
\rangle^2 }{(-1)^{1+3q} 2^{1+q}\beta q (2q-1) }\right]^n  \chi (z),
\end{eqnarray}
with
\begin{eqnarray}\label{mn}
n=1,~~~m=3-2q.
\end{eqnarray}
Thus, Eq. (\ref{phi}) becomes
\begin{eqnarray}\label{phix}
\chi^{\prime\prime}(z)-\frac{1}{z}
\left(\frac{d-2}{2q-1}-2\right)\chi^\prime(z)
=\frac{z^{2[\lambda_i- q-1+(\frac{d-2}{2q-1}-1)(1-q)]}\left(1-z^{\frac{d-2}{2q-1}-1}\right)F^2(z)}{g(z)},
\end{eqnarray}
which shows us that the functional $\chi(z)$ is independent on
$r_+$, $T_c$ and $\langle\mathcal{O}_i \rangle$. For examples: if we
take $d=5$ and $q=3/4$, from Eq. (\ref{phix}) we can easily find
that  $ \chi(z)=c_1+c_2 z^5+ \Big[-\frac{a^2 z^8}{4 \lambda_i^2+22
\lambda_i+24}+a \left(\frac{z}{2 \lambda_i^2+7
\lambda_i+3}+\frac{1}{2 \lambda_i^2+5 \lambda_i}\right)
z^5+\frac{z^3}{-4 \lambda_i^2-2 \lambda_i+6}\Big] z^{2
\lambda_i}$; and if we take $d=5$ and $q=1$, we have $ \chi(z)=
c_1+c_2 z^2+\frac{1}{4
\lambda_i\left(\lambda_i^2-1\right)}\Big[-(\lambda_i+1)
z^{2\lambda_i}+(2 a+1) (\lambda_i-1) z^{2
\lambda_i+2}-(a+1)^2(\lambda_i-1)\lambda_i
\Gamma(\lambda_i+2)\,_2\tilde{F}_1\left(\lambda_i+1,1;
\lambda_i+3;-z^2\right) z^{2\lambda_i+4}\Big].$ That is to say, Eq.
(\ref{phix}) tells us that $\chi (z)|_{z=0}=c_1$ is a constant which
is also independent on $r_+$, $T_c$ and $\langle\mathcal{O}_i
\rangle$.

At the boundary $z=0$, from Eqs. (\ref{phin}) and (\ref{phis}) we have
\begin{eqnarray}\label{rel1}
A \frac{T_c^{\frac{d-2}{2q-1}}} {T^{\frac{d-2}{2q-1}-1}}-(A T_c)
=\left[\frac{r_+^{2 q-2\lambda_i-2} \langle\mathcal{O}_i \rangle^2
}{(-1)^{1+3q}2^{1+q}\beta q (2q-1) }\right] (A T_c)^{3-2q} c_1.
\end{eqnarray}
After some calculations, Eq. (\ref{rel1}) can be casted into
\begin{eqnarray}\label{rel2}
 \frac{\langle\mathcal{O}_i \rangle}{T_c^{\lambda_i}}=
 D \left(\frac{T}{T_c}\right)^{\lambda_i+(1-q)}
 \left\{ \left(\frac{T_c}{ T}\right)^{\frac{d-2}{2q-1}-1}
 \left[1-\left(\frac{T}{T_c} \right)^{\frac{d-2}{2q-1}-1}
 \right]\right\}^{\frac{1}{2}},
\end{eqnarray}
where constant $D$ is independent on $\langle\mathcal{O}_i \rangle$,
$T$ and $T_c$. Note that our result (\ref{rel2}) is valid for both
of the scalar operators $\langle\mathcal{O}_+ \rangle$ and
$\langle\mathcal{O}_- \rangle$ with various Power-Maxwell parameters
$q$ and mass $m$ of the scalar field. It is interesting to point out
that the critical exponent of the system is equal to $1/2$ which is
in agreement with the mean field value.
\FIGURE{
\includegraphics[scale=0.68]{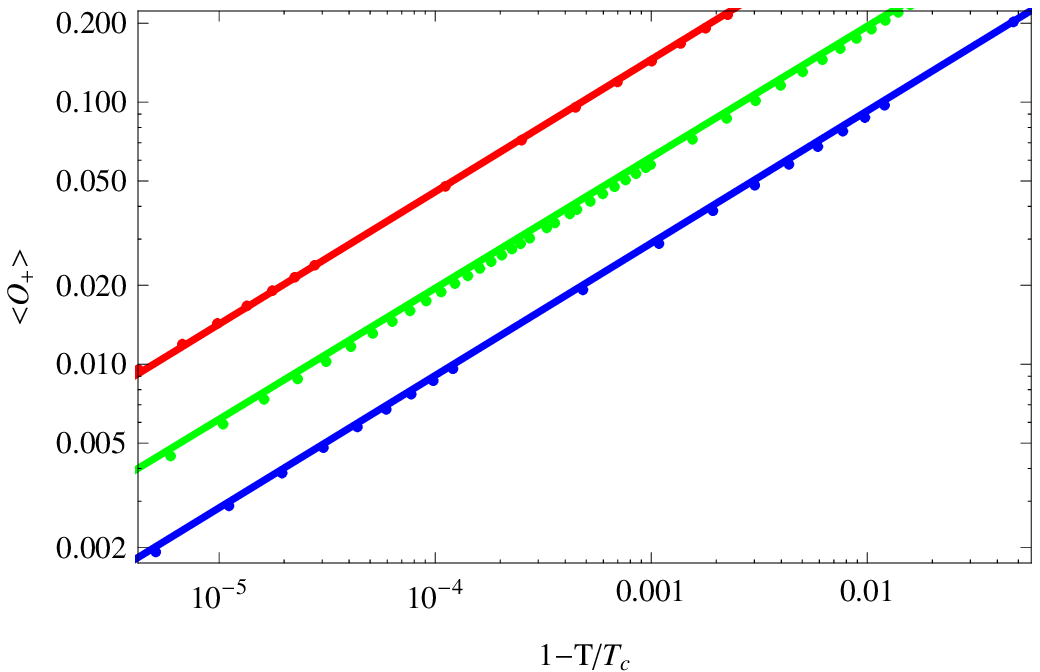}\hspace{0.2cm}%
\includegraphics[scale=0.68]{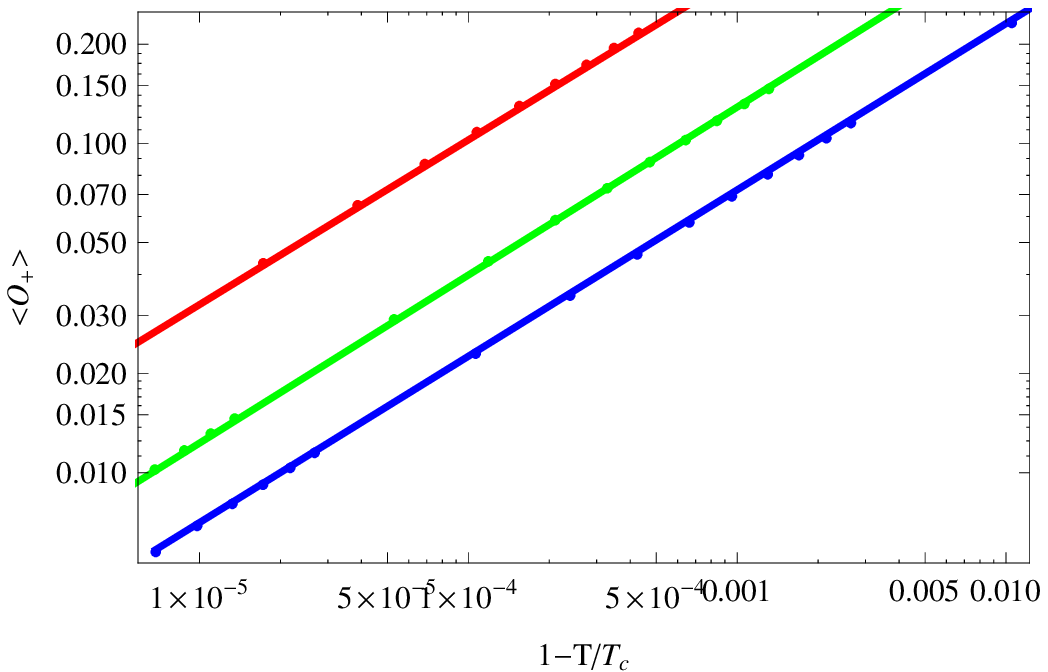}\hspace{0.2cm}%
\caption{\label{Fig1} (color online) The condensate
$\langle\mathcal{O}_+ \rangle$ vs $1-T/T_{c}$ in logarithmic scale
with different values of $q$ for $d=4$ (left) and $d=5$ (right). The
three lines  from bottom to top in left panel correspond to $q=0.75$
(blue), $0.85$ (green) and $1.0$ (red), and in right one correspond
to $q=0.875$ (blue), $1.0$ (green) and $1.25$ (red). These panels
show that the slope is independent of $q$.}
}

To check the analytical result (\ref{rel2}) obtained by using the analytical
method, we calculate the critical exponent of the system by
using numerical approach. In Fig. \ref{Fig1}, we present the
condensate $\langle\mathcal{O}_+ \rangle$ as a function of
$(1-T/T_{c})$ in logarithmic scale with different values of $q$ for
$d=4$ ( left) and $d=5$ (right). The three lines from bottom to top
in left panel correspond to $q=0.75$, $0.85$  and $1.0$, and in
right one correspond to $q=0.875$, $1.0$ and $1.25$. We see  from
these panels that the slope is almost independent of the power parameter
$q$ and the mass $m$ of the scalar field, which is in agreement with
the analytical value $1/2$.

\section{conclusions}

The behaviors of the holographic superconductors have been
investigated by introducing a complex charged scalar field coupled
with a Power-Maxwell field in the background of a planar
Schwarzschild AdS black hole. We present a detail analysis of the
condensation of the operators by using both the Sturm-Liouville
analytical and numerical methods. We first note that the Power-Maxwell field takes the special asymptotical solution near the boundary which is different from all known cases. It is interesting to find that, if we fix the mass parameter $m$, the critical temperature decreases as the $q$ increases, which means that the larger power parameter $q$
for the Power-Maxwell field makes it harder for the scalar hair to
be condensated. For the same $q$, the critical temperature decreases as the value of $m^2$ increases, which means that the scalar hair
can be formed more difficult for the larger mass of the scalar field in both 4-dimensional and 5-dimensional Schwarzschild AdS black
holes, which agrees with the finding in Ref. \cite{HorowitzPRD78} when $q=1$. We finally find that, for both the scalar operators
$\langle\mathcal{O}_+ \rangle$ and $\langle\mathcal{O}_- \rangle$
with different power parameters $q$ and masses $m$, the critical
exponent of the system is always $1/2$, which seems to be an universal property for various nonlinear electrodynamics if the scalar field $\psi$ takes the form of this paper.

\begin{acknowledgments}

This work was supported by the National Natural Science Foundation of China under Grant Nos. 11175065, 10935013; the National Basic Research of China under Grant No. 2010CB833004; PCSIRT, No. IRT0964; the Hunan Provincial Natural Science Foundation of China under Grant No. 11JJ7001; and Construct Program of the National  Key Discipline. We thank the Kavli Institute for Theoretical Physics China for hospitality in the revised stages of this work. 	

\end{acknowledgments}

\newpage
\appendix
\addcontentsline{toc}{section}{Appendix}
\section*{Appendix}
\section{ $\phi$ behavior at the asymptotic AdS region}

In the AdS/CFT duality \cite{polyakov,Witten}, it is well known that the bulk gauge field $A_\mu$ acts as a source for a conserved  current $J^\mu$ corresponding to a global U(1) symmetry. In general, near the boundary, the equation of motion for $A_\mu$ has the solution
\begin{eqnarray}
A_\mu =a_\mu - \frac{b_\mu}{r^c} +\ldots ,
\label{ABoundary}
\end{eqnarray}
where $a_\mu$ and $b_\mu$ are constant which physical properties will be discussed in following.
Given this expansion, the one-point function of the corresponding field theory operator can be expressed as variation of the action with respect to the boundary value \cite{HerzogRev}
\begin{eqnarray}
\langle J^\mu \rangle \sim  \frac{\delta S}{\delta a_\mu} .
\label{onepoint}
\end{eqnarray}
From the fact that $\langle J^\mu \rangle$ and $a_\mu$ are
canonically conjugate, we can identify $a_0$ with the
chemical potential $\mu$. Then, the charge density $\rho$ in field theory is defined as \cite{HerzogRev}
\begin{eqnarray}
\rho=\langle J^{0} \rangle .
\label{CDensity}
\end{eqnarray}

For the system that the Power-Maxwell field is coupled with the charged scalar field, we have the solution $A_0=\phi=a_0-b_0/ r^{\frac{d-2}{2q-1}-1}$ near the boundary. Thus, on shell, i.e. evaluated for a solution to the classical equations of motion, the action reduces to a boundary term of the form
\begin{eqnarray}
S=\left[B a_0 b_0^{2q-1}+(d-1)\Psi_+\Psi_1+\lambda_-r_B^{d-1-2\lambda_-} \Psi_-^2\right] \cdot \int d^{d-1}x \sqrt{-g_{d-1}}+\ldots ,
\label{Action}
\end{eqnarray}
where $B=-(-2)^q\beta \left(\frac{d-2}{2q-1}\right)^{2q-1}$. Using Eqs. (\ref{onepoint}), (\ref{CDensity}) and (\ref{Action}), we know that the charge density is given by
\begin{eqnarray}
\rho=b_0^{2q-1}.
\label{CDensity1}
\end{eqnarray}
Therefore, at the asymptotic AdS region
($r\rightarrow\infty$), the solution behave for function $\phi$ is
\begin{eqnarray}
\phi=\mu-\frac{\rho^\frac{1}{2q-1}}{r^{\frac{d-2}{2q-1}-1}}. \label{infinity1}
\end{eqnarray}
We should point out that the electric field $\phi$ is dependent on the coupling constant $q$ at the asymptotic AdS region, which in different from all known cases.

\end{document}